\documentclass[hyperindex,breaklinks=true, colorlinks, citecolor=blue]{aa}  
\usepackage{graphicx}
\usepackage{epstopdf}

\usepackage{txfonts}
\usepackage{amssymb}
\usepackage{aas_macros}
\usepackage{subeqnarray}
\usepackage{natbib}
\usepackage{version}

\usepackage{color}
\usepackage{txfonts}
\usepackage[mathscr]{euscript}

\def\setsymbol#1#2{\expandafter\def\csname #1\endcsname{#2}}
\def\getsymbol#1{\csname #1\endcsname}

\def\Planck{\textit{Planck}}

\def\all2013resultspapers{\nocite{planck2013-p01, planck2013-p02, planck2013-p02a, planck2013-p02d, planck2013-p02b, planck2013-p03, planck2013-p03c, planck2013-p03f, planck2013-p03d, planck2013-p03e, planck2013-p01a, planck2013-p06, planck2013-p03a, planck2013-pip88, planck2013-p08, planck2013-p11, planck2013-p12, planck2013-p13, planck2013-p14, planck2013-p15, planck2013-p05b, planck2013-p17, planck2013-p09, planck2013-p09a, planck2013-p20, planck2013-p19, planck2013-pipaberration, planck2013-p05, planck2013-p05a, planck2013-pip56, planck2013-p06b}}

\newbox\tablebox    \newdimen\tablewidth
\def\leaderfil{\leaders\hbox to 5pt{\hss.\hss}\hfil}
\def\tablenote#1 #2\par{\begingroup \parindent=0.8em
    \abovedisplayshortskip=0pt\belowdisplayshortskip=0pt
    \noindent
    $$\hss\vbox{\hsize\tablewidth \hangindent=\parindent \hangafter=1 \noindent
    \hbox to \parindent{$^#1$\hss}\strut#2\strut\par}\hss$$
    \endgroup}

\def\L2{\ifmmode L_2\else $L_2$\fi}

\def\DeltaT{\ifmmode \Delta T\else $\Delta T$\fi}
\def\deltat{\ifmmode \Delta t\else $\Delta t$\fi}
\def\fknee{\ifmmode f_{\rm knee}\else $f_{\rm knee}$\fi}
\def\Fmax{\ifmmode F_{\rm max}\else $F_{\rm max}$\fi}
\def\solar{\ifmmode{\rm M}_{\mathord\odot}\else${\rm M}_{\mathord\odot}$\fi}
\def\Msolar{\ifmmode{\rm M}_{\mathord\odot}\else${\rm M}_{\mathord\odot}$\fi}
\def\Lsolar{\ifmmode{\rm L}_{\mathord\odot}\else${\rm L}_{\mathord\odot}$\fi}

\def\inv{\ifmmode^{-1}\else$^{-1}$\fi}
\def\mo{\ifmmode^{-1}\else$^{-1}$\fi}
\def\sup#1{\ifmmode ^{\rm #1}\else $^{\rm #1}$\fi}
\def\expo#1{\ifmmode \times 10^{#1}\else $\times 10^{#1}$\fi}
\def\,{\thinspace}
\def\lsim{\mathrel{\raise .4ex\hbox{\rlap{$<$}\lower 1.2ex\hbox{$\sim$}}}}
\def\gsim{\mathrel{\raise .4ex\hbox{\rlap{$>$}\lower 1.2ex\hbox{$\sim$}}}}

\def\simprop{\mathrel{\raise .4ex\hbox{\rlap{$\propto$}\lower 1.2ex\hbox{$\sim$}}}}
\def\deg{\ifmmode^\circ\else$^\circ$\fi}
\def\pdeg{\ifmmode $\setbox0=\hbox{$^{\circ}$}\rlap{\hskip.11\wd0 .}$^{\circ}
          \else \setbox0=\hbox{$^{\circ}$}\rlap{\hskip.11\wd0 .}$^{\circ}$\fi}
\def\arcs{\ifmmode {^{\scriptstyle\prime\prime}}
          \else $^{\scriptstyle\prime\prime}$\fi}
\def\arcm{\ifmmode {^{\scriptstyle\prime}}
          \else $^{\scriptstyle\prime}$\fi}
\newdimen\sa  \newdimen\sb
\def\parcs{\sa=.07em \sb=.03em
     \ifmmode \hbox{\rlap{.}}^{\scriptstyle\prime\kern -\sb\prime}\hbox{\kern -\sa}
     \else \rlap{.}$^{\scriptstyle\prime\kern -\sb\prime}$\kern -\sa\fi}
\def\parcm{\sa=.08em \sb=.03em
     \ifmmode \hbox{\rlap{.}\kern\sa}^{\scriptstyle\prime}\hbox{\kern-\sb}
     \else \rlap{.}\kern\sa$^{\scriptstyle\prime}$\kern-\sb\fi}
\def\ra[#1 #2 #3.#4]{#1\sup{h}#2\sup{m}#3\sup{s}\llap.#4}
\def\dec[#1 #2 #3.#4]{#1\deg#2\arcm#3\arcs\llap.#4}
\def\deco[#1 #2 #3]{#1\deg#2\arcm#3\arcs}
\def\rra[#1 #2]{#1\sup{h}#2\sup{m}}

\def\dots{\relax\ifmmode \ldots\else $\ldots$\fi}
\def\WHzsr{\ifmmode $W\,Hz\mo\,sr\mo$\else W\,Hz\mo\,sr\mo\fi}
\def\mHz{\ifmmode $\,mHz$\else \,mHz\fi}
\def\GHz{\ifmmode $\,GHz$\else \,GHz\fi}
\def\mKs{\ifmmode $\,mK\,s$^{1/2}\else \,mK\,s$^{1/2}$\fi}
\def\muKs{\ifmmode \,\mu$K\,s$^{1/2}\else \,$\mu$K\,s$^{1/2}$\fi}
\def\muKRJs{\ifmmode \,\mu$K$_{\rm RJ}$\,s$^{1/2}\else \,$\mu$K$_{\rm RJ}$\,s$^{1/2}$\fi}
\def\muKHz{\ifmmode \,\mu$K\,Hz$^{-1/2}\else \,$\mu$K\,Hz$^{-1/2}$\fi}
\def\MJysr{\ifmmode \,$MJy\,sr\mo$\else \,MJy\,sr\mo\fi}
\def\MJysrmK{\ifmmode \,$MJy\,sr\mo$\,mK$_{\rm CMB}\mo\else \,MJy\,sr\mo\,mK$_{\rm CMB}\mo$\fi}
\def\microns{\ifmmode \,\mu$m$\else \,$\mu$m\fi}

\def\muK{\ifmmode \,\mu$K$\else \,$\mu$\hbox{K}\fi}
\def\microK{\ifmmode \,\mu$K$\else \,$\mu$\hbox{K}\fi}
\def\muW{\ifmmode \,\mu$W$\else \,$\mu$\hbox{W}\fi}
\def\kms{\ifmmode $\,km\,s$^{-1}\else \,km\,s$^{-1}$\fi}
\def\kmsMpc{\ifmmode $\,\kms\,Mpc\mo$\else \,\kms\,Mpc\mo\fi}
\setsymbol{LFI:center:frequency:70GHz:units}{70.3\,GHz}
\setsymbol{LFI:center:frequency:44GHz:units}{44.1\,GHz}
\setsymbol{LFI:center:frequency:30GHz:units}{28.5\,GHz}

\setsymbol{LFI:center:frequency:70GHz}{70.3}
\setsymbol{LFI:center:frequency:44GHz}{44.1}
\setsymbol{LFI:center:frequency:30GHz}{28.5}

\setsymbol{LFI:center:frequency:LFI18:Rad:M:units}{71.7\GHz}
\setsymbol{LFI:center:frequency:LFI19:Rad:M:units}{67.5\GHz}
\setsymbol{LFI:center:frequency:LFI20:Rad:M:units}{69.2\GHz}
\setsymbol{LFI:center:frequency:LFI21:Rad:M:units}{70.4\GHz}
\setsymbol{LFI:center:frequency:LFI22:Rad:M:units}{71.5\GHz}
\setsymbol{LFI:center:frequency:LFI23:Rad:M:units}{70.8\GHz}
\setsymbol{LFI:center:frequency:LFI24:Rad:M:units}{44.4\GHz}
\setsymbol{LFI:center:frequency:LFI25:Rad:M:units}{44.0\GHz}
\setsymbol{LFI:center:frequency:LFI26:Rad:M:units}{43.9\GHz}
\setsymbol{LFI:center:frequency:LFI27:Rad:M:units}{28.3\GHz}
\setsymbol{LFI:center:frequency:LFI28:Rad:M:units}{28.8\GHz}
\setsymbol{LFI:center:frequency:LFI18:Rad:S:units}{70.1\GHz}
\setsymbol{LFI:center:frequency:LFI19:Rad:S:units}{69.6\GHz}
\setsymbol{LFI:center:frequency:LFI20:Rad:S:units}{69.5\GHz}
\setsymbol{LFI:center:frequency:LFI21:Rad:S:units}{69.5\GHz}
\setsymbol{LFI:center:frequency:LFI22:Rad:S:units}{72.8\GHz}
\setsymbol{LFI:center:frequency:LFI23:Rad:S:units}{71.3\GHz}
\setsymbol{LFI:center:frequency:LFI24:Rad:S:units}{44.1\GHz}
\setsymbol{LFI:center:frequency:LFI25:Rad:S:units}{44.1\GHz}
\setsymbol{LFI:center:frequency:LFI26:Rad:S:units}{44.1\GHz}
\setsymbol{LFI:center:frequency:LFI27:Rad:S:units}{28.5\GHz}
\setsymbol{LFI:center:frequency:LFI28:Rad:S:units}{28.2\GHz}

\setsymbol{LFI:center:frequency:LFI18:Rad:M}{71.7}
\setsymbol{LFI:center:frequency:LFI19:Rad:M}{67.5}
\setsymbol{LFI:center:frequency:LFI20:Rad:M}{69.2}
\setsymbol{LFI:center:frequency:LFI21:Rad:M}{70.4}
\setsymbol{LFI:center:frequency:LFI22:Rad:M}{71.5}
\setsymbol{LFI:center:frequency:LFI23:Rad:M}{70.8}
\setsymbol{LFI:center:frequency:LFI24:Rad:M}{44.4}
\setsymbol{LFI:center:frequency:LFI25:Rad:M}{44.0}
\setsymbol{LFI:center:frequency:LFI26:Rad:M}{43.9}
\setsymbol{LFI:center:frequency:LFI27:Rad:M}{28.3}
\setsymbol{LFI:center:frequency:LFI28:Rad:M}{28.8}
\setsymbol{LFI:center:frequency:LFI18:Rad:S}{70.1}
\setsymbol{LFI:center:frequency:LFI19:Rad:S}{69.6}
\setsymbol{LFI:center:frequency:LFI20:Rad:S}{69.5}
\setsymbol{LFI:center:frequency:LFI21:Rad:S}{69.5}
\setsymbol{LFI:center:frequency:LFI22:Rad:S}{72.8}
\setsymbol{LFI:center:frequency:LFI23:Rad:S}{71.3}
\setsymbol{LFI:center:frequency:LFI24:Rad:S}{44.1}
\setsymbol{LFI:center:frequency:LFI25:Rad:S}{44.1}
\setsymbol{LFI:center:frequency:LFI26:Rad:S}{44.1}
\setsymbol{LFI:center:frequency:LFI27:Rad:S}{28.5}
\setsymbol{LFI:center:frequency:LFI28:Rad:S}{28.2}

\setsymbol{LFI:white:noise:sensitivity:70GHz:units}{134.7\muKs}
\setsymbol{LFI:white:noise:sensitivity:44GHz:units}{164.7\muKs}
\setsymbol{LFI:white:noise:sensitivity:30GHz:units}{143.4\muKs}

\setsymbol{LFI:white:noise:sensitivity:70GHz}{134.7}
\setsymbol{LFI:white:noise:sensitivity:44GHz}{164.7}
\setsymbol{LFI:white:noise:sensitivity:30GHz}{143.4}

\setsymbol{LFI:white:noise:sensitivity:LFI18:Rad:M:units}{512.0\muKs}
\setsymbol{LFI:white:noise:sensitivity:LFI19:Rad:M:units}{581.4\muKs}
\setsymbol{LFI:white:noise:sensitivity:LFI20:Rad:M:units}{590.8\muKs}
\setsymbol{LFI:white:noise:sensitivity:LFI21:Rad:M:units}{455.2\muKs}
\setsymbol{LFI:white:noise:sensitivity:LFI22:Rad:M:units}{492.0\muKs}
\setsymbol{LFI:white:noise:sensitivity:LFI23:Rad:M:units}{507.7\muKs}
\setsymbol{LFI:white:noise:sensitivity:LFI24:Rad:M:units}{462.2\muKs}
\setsymbol{LFI:white:noise:sensitivity:LFI25:Rad:M:units}{413.6\muKs}
\setsymbol{LFI:white:noise:sensitivity:LFI26:Rad:M:units}{478.6\muKs}
\setsymbol{LFI:white:noise:sensitivity:LFI27:Rad:M:units}{277.7\muKs}
\setsymbol{LFI:white:noise:sensitivity:LFI28:Rad:M:units}{312.3\muKs}
\setsymbol{LFI:white:noise:sensitivity:LFI18:Rad:S:units}{465.7\muKs}
\setsymbol{LFI:white:noise:sensitivity:LFI19:Rad:S:units}{555.6\muKs}
\setsymbol{LFI:white:noise:sensitivity:LFI20:Rad:S:units}{623.2\muKs}
\setsymbol{LFI:white:noise:sensitivity:LFI21:Rad:S:units}{564.1\muKs}
\setsymbol{LFI:white:noise:sensitivity:LFI22:Rad:S:units}{534.4\muKs}
\setsymbol{LFI:white:noise:sensitivity:LFI23:Rad:S:units}{542.4\muKs}
\setsymbol{LFI:white:noise:sensitivity:LFI24:Rad:S:units}{399.2\muKs}
\setsymbol{LFI:white:noise:sensitivity:LFI25:Rad:S:units}{392.6\muKs}
\setsymbol{LFI:white:noise:sensitivity:LFI26:Rad:S:units}{418.6\muKs}
\setsymbol{LFI:white:noise:sensitivity:LFI27:Rad:S:units}{302.9\muKs}
\setsymbol{LFI:white:noise:sensitivity:LFI28:Rad:S:units}{285.3\muKs}

\setsymbol{LFI:white:noise:sensitivity:LFI18:Rad:M}{512.0}
\setsymbol{LFI:white:noise:sensitivity:LFI19:Rad:M}{581.4}
\setsymbol{LFI:white:noise:sensitivity:LFI20:Rad:M}{590.8}
\setsymbol{LFI:white:noise:sensitivity:LFI21:Rad:M}{455.2}
\setsymbol{LFI:white:noise:sensitivity:LFI22:Rad:M}{492.0}
\setsymbol{LFI:white:noise:sensitivity:LFI23:Rad:M}{507.7}
\setsymbol{LFI:white:noise:sensitivity:LFI24:Rad:M}{462.2}
\setsymbol{LFI:white:noise:sensitivity:LFI25:Rad:M}{413.6}
\setsymbol{LFI:white:noise:sensitivity:LFI26:Rad:M}{478.6}
\setsymbol{LFI:white:noise:sensitivity:LFI27:Rad:M}{277.7}
\setsymbol{LFI:white:noise:sensitivity:LFI28:Rad:M}{312.3}
\setsymbol{LFI:white:noise:sensitivity:LFI18:Rad:S}{465.7}
\setsymbol{LFI:white:noise:sensitivity:LFI19:Rad:S}{555.6}
\setsymbol{LFI:white:noise:sensitivity:LFI20:Rad:S}{623.2}
\setsymbol{LFI:white:noise:sensitivity:LFI21:Rad:S}{564.1}
\setsymbol{LFI:white:noise:sensitivity:LFI22:Rad:S}{534.4}
\setsymbol{LFI:white:noise:sensitivity:LFI23:Rad:S}{542.4}
\setsymbol{LFI:white:noise:sensitivity:LFI24:Rad:S}{399.2}
\setsymbol{LFI:white:noise:sensitivity:LFI25:Rad:S}{392.6}
\setsymbol{LFI:white:noise:sensitivity:LFI26:Rad:S}{418.6}
\setsymbol{LFI:white:noise:sensitivity:LFI27:Rad:S}{302.9}
\setsymbol{LFI:white:noise:sensitivity:LFI28:Rad:S}{285.3}

\setsymbol{LFI:knee:frequency:70GHz:units}{29.5\mHz}
\setsymbol{LFI:knee:frequency:44GHz:units}{56.2\mHz}
\setsymbol{LFI:knee:frequency:30GHz:units}{113.7\mHz}

\setsymbol{LFI:knee:frequency:70GHz}{29.5}
\setsymbol{LFI:knee:frequency:44GHz}{56.2}
\setsymbol{LFI:knee:frequency:30GHz}{113.7}

\setsymbol{LFI:knee:frequency:LFI18:Rad:M:units}{16.3\mHz}
\setsymbol{LFI:knee:frequency:LFI19:Rad:M:units}{15.1\mHz}
\setsymbol{LFI:knee:frequency:LFI20:Rad:M:units}{18.7\mHz}
\setsymbol{LFI:knee:frequency:LFI21:Rad:M:units}{37.2\mHz}
\setsymbol{LFI:knee:frequency:LFI22:Rad:M:units}{12.7\mHz}
\setsymbol{LFI:knee:frequency:LFI23:Rad:M:units}{34.6\mHz}
\setsymbol{LFI:knee:frequency:LFI24:Rad:M:units}{46.2\mHz}
\setsymbol{LFI:knee:frequency:LFI25:Rad:M:units}{24.9\mHz}
\setsymbol{LFI:knee:frequency:LFI26:Rad:M:units}{67.6\mHz}
\setsymbol{LFI:knee:frequency:LFI27:Rad:M:units}{187.4\mHz}
\setsymbol{LFI:knee:frequency:LFI28:Rad:M:units}{122.2\mHz}
\setsymbol{LFI:knee:frequency:LFI18:Rad:S:units}{17.7\mHz}
\setsymbol{LFI:knee:frequency:LFI19:Rad:S:units}{22.0\mHz}
\setsymbol{LFI:knee:frequency:LFI20:Rad:S:units}{8.7\mHz}
\setsymbol{LFI:knee:frequency:LFI21:Rad:S:units}{25.9\mHz}
\setsymbol{LFI:knee:frequency:LFI22:Rad:S:units}{15.8\mHz}
\setsymbol{LFI:knee:frequency:LFI23:Rad:S:units}{129.8\mHz}
\setsymbol{LFI:knee:frequency:LFI24:Rad:S:units}{100.9\mHz}
\setsymbol{LFI:knee:frequency:LFI25:Rad:S:units}{38.9\mHz}
\setsymbol{LFI:knee:frequency:LFI26:Rad:S:units}{58.9\mHz}
\setsymbol{LFI:knee:frequency:LFI27:Rad:S:units}{104.4\mHz}
\setsymbol{LFI:knee:frequency:LFI28:Rad:S:units}{40.7\mHz}

\setsymbol{LFI:knee:frequency:LFI18:Rad:M}{16.3}
\setsymbol{LFI:knee:frequency:LFI19:Rad:M}{15.1}
\setsymbol{LFI:knee:frequency:LFI20:Rad:M}{18.7}
\setsymbol{LFI:knee:frequency:LFI21:Rad:M}{37.2}
\setsymbol{LFI:knee:frequency:LFI22:Rad:M}{12.7}
\setsymbol{LFI:knee:frequency:LFI23:Rad:M}{34.6}
\setsymbol{LFI:knee:frequency:LFI24:Rad:M}{46.2}
\setsymbol{LFI:knee:frequency:LFI25:Rad:M}{24.9}
\setsymbol{LFI:knee:frequency:LFI26:Rad:M}{67.6}
\setsymbol{LFI:knee:frequency:LFI27:Rad:M}{187.4}
\setsymbol{LFI:knee:frequency:LFI28:Rad:M}{122.2}
\setsymbol{LFI:knee:frequency:LFI18:Rad:S}{17.7}
\setsymbol{LFI:knee:frequency:LFI19:Rad:S}{22.0}
\setsymbol{LFI:knee:frequency:LFI20:Rad:S}{8.7}
\setsymbol{LFI:knee:frequency:LFI21:Rad:S}{25.9}
\setsymbol{LFI:knee:frequency:LFI22:Rad:S}{15.8}
\setsymbol{LFI:knee:frequency:LFI23:Rad:S}{129.8}
\setsymbol{LFI:knee:frequency:LFI24:Rad:S}{100.9}
\setsymbol{LFI:knee:frequency:LFI25:Rad:S}{38.9}
\setsymbol{LFI:knee:frequency:LFI26:Rad:S}{58.9}
\setsymbol{LFI:knee:frequency:LFI27:Rad:S}{104.4}
\setsymbol{LFI:knee:frequency:LFI28:Rad:S}{40.7}

\setsymbol{LFI:slope:70GHz:units}{$-1.03$\mHz}
\setsymbol{LFI:slope:44GHz:units}{$-0.89$\mHz}
\setsymbol{LFI:slope:30GHz:units}{$-0.87$\mHz}

\setsymbol{LFI:slope:70GHz}{$-1.03$}
\setsymbol{LFI:slope:44GHz}{$-0.89$}
\setsymbol{LFI:slope:30GHz}{$-0.87$}

\setsymbol{LFI:slope:LFI18:Rad:M:units}{$-1.04$\mHz}
\setsymbol{LFI:slope:LFI19:Rad:M:units}{$-1.09$\mHz}
\setsymbol{LFI:slope:LFI20:Rad:M:units}{$-0.69$\mHz}
\setsymbol{LFI:slope:LFI21:Rad:M:units}{$-1.56$\mHz}
\setsymbol{LFI:slope:LFI22:Rad:M:units}{$-1.01$\mHz}
\setsymbol{LFI:slope:LFI23:Rad:M:units}{$-0.96$\mHz}
\setsymbol{LFI:slope:LFI24:Rad:M:units}{$-0.83$\mHz}
\setsymbol{LFI:slope:LFI25:Rad:M:units}{$-0.91$\mHz}
\setsymbol{LFI:slope:LFI26:Rad:M:units}{$-0.95$\mHz}
\setsymbol{LFI:slope:LFI27:Rad:M:units}{$-0.87$\mHz}
\setsymbol{LFI:slope:LFI28:Rad:M:units}{$-0.88$\mHz}
\setsymbol{LFI:slope:LFI18:Rad:S:units}{$-1.15$\mHz}
\setsymbol{LFI:slope:LFI19:Rad:S:units}{$-1.00$\mHz}
\setsymbol{LFI:slope:LFI20:Rad:S:units}{$-0.95$\mHz}
\setsymbol{LFI:slope:LFI21:Rad:S:units}{$-0.92$\mHz}
\setsymbol{LFI:slope:LFI22:Rad:S:units}{$-1.01$\mHz}
\setsymbol{LFI:slope:LFI23:Rad:S:units}{$-0.95$\mHz}
\setsymbol{LFI:slope:LFI24:Rad:S:units}{$-0.73$\mHz}
\setsymbol{LFI:slope:LFI25:Rad:S:units}{$-1.16$\mHz}
\setsymbol{LFI:slope:LFI26:Rad:S:units}{$-0.79$\mHz}
\setsymbol{LFI:slope:LFI27:Rad:S:units}{$-0.82$\mHz}
\setsymbol{LFI:slope:LFI28:Rad:S:units}{$-0.91$\mHz}

\setsymbol{LFI:slope:LFI18:Rad:M}{$-1.04$}
\setsymbol{LFI:slope:LFI19:Rad:M}{$-1.09$}
\setsymbol{LFI:slope:LFI20:Rad:M}{$-0.69$}
\setsymbol{LFI:slope:LFI21:Rad:M}{$-1.56$}
\setsymbol{LFI:slope:LFI22:Rad:M}{$-1.01$}
\setsymbol{LFI:slope:LFI23:Rad:M}{$-0.96$}
\setsymbol{LFI:slope:LFI24:Rad:M}{$-0.83$}
\setsymbol{LFI:slope:LFI25:Rad:M}{$-0.91$}
\setsymbol{LFI:slope:LFI26:Rad:M}{$-0.95$}
\setsymbol{LFI:slope:LFI27:Rad:M}{$-0.87$}
\setsymbol{LFI:slope:LFI28:Rad:M}{$-0.88$}
\setsymbol{LFI:slope:LFI18:Rad:S}{$-1.15$}
\setsymbol{LFI:slope:LFI19:Rad:S}{$-1.00$}
\setsymbol{LFI:slope:LFI20:Rad:S}{$-0.95$}
\setsymbol{LFI:slope:LFI21:Rad:S}{$-0.92$}
\setsymbol{LFI:slope:LFI22:Rad:S}{$-1.01$}
\setsymbol{LFI:slope:LFI23:Rad:S}{$-0.95$}
\setsymbol{LFI:slope:LFI24:Rad:S}{$-0.73$}
\setsymbol{LFI:slope:LFI25:Rad:S}{$-1.16$}
\setsymbol{LFI:slope:LFI26:Rad:S}{$-0.79$}
\setsymbol{LFI:slope:LFI27:Rad:S}{$-0.82$}
\setsymbol{LFI:slope:LFI28:Rad:S}{$-0.91$}

\setsymbol{LFI:FWHM:70GHz:units}{13\parcm01}
\setsymbol{LFI:FWHM:44GHz:units}{27\parcm92}
\setsymbol{LFI:FWHM:30GHz:units}{32\parcm65}

\setsymbol{LFI:FWHM:70GHz}{13.01}
\setsymbol{LFI:FWHM:44GHz}{27.92}
\setsymbol{LFI:FWHM:30GHz}{32.65}

\setsymbol{LFI:FWHM:LFI18:units}{13\parcm39}
\setsymbol{LFI:FWHM:LFI19:units}{13\parcm01}
\setsymbol{LFI:FWHM:LFI20:units}{12\parcm75}
\setsymbol{LFI:FWHM:LFI21:units}{12\parcm74}
\setsymbol{LFI:FWHM:LFI22:units}{12\parcm87}
\setsymbol{LFI:FWHM:LFI23:units}{13\parcm27}
\setsymbol{LFI:FWHM:LFI24:units}{22\parcm98}
\setsymbol{LFI:FWHM:LFI25:units}{30\parcm46}
\setsymbol{LFI:FWHM:LFI26:units}{30\parcm31}
\setsymbol{LFI:FWHM:LFI27:units}{32\parcm65}
\setsymbol{LFI:FWHM:LFI28:units}{32\parcm66}

\setsymbol{LFI:FWHM:LFI18}{13.39}
\setsymbol{LFI:FWHM:LFI19}{13.01}
\setsymbol{LFI:FWHM:LFI20}{12.75}
\setsymbol{LFI:FWHM:LFI21}{12.74}
\setsymbol{LFI:FWHM:LFI22}{12.87}
\setsymbol{LFI:FWHM:LFI23}{13.27}
\setsymbol{LFI:FWHM:LFI24}{22.98}
\setsymbol{LFI:FWHM:LFI25}{30.46}
\setsymbol{LFI:FWHM:LFI26}{30.31}
\setsymbol{LFI:FWHM:LFI27}{32.65}
\setsymbol{LFI:FWHM:LFI28}{32.66}

\setsymbol{LFI:FWHM:uncertainty:LFI18:units}{0.170\arcm}
\setsymbol{LFI:FWHM:uncertainty:LFI19:units}{0.174\arcm}
\setsymbol{LFI:FWHM:uncertainty:LFI20:units}{0.170\arcm}
\setsymbol{LFI:FWHM:uncertainty:LFI21:units}{0.156\arcm}
\setsymbol{LFI:FWHM:uncertainty:LFI22:units}{0.164\arcm}
\setsymbol{LFI:FWHM:uncertainty:LFI23:units}{0.171\arcm}
\setsymbol{LFI:FWHM:uncertainty:LFI24:units}{0.652\arcm}
\setsymbol{LFI:FWHM:uncertainty:LFI25:units}{1.075\arcm}
\setsymbol{LFI:FWHM:uncertainty:LFI26:units}{1.131\arcm}
\setsymbol{LFI:FWHM:uncertainty:LFI27:units}{1.266\arcm}
\setsymbol{LFI:FWHM:uncertainty:LFI28:units}{1.287\arcm}

\setsymbol{LFI:FWHM:uncertainty:LFI18}{0.170}
\setsymbol{LFI:FWHM:uncertainty:LFI19}{0.174}
\setsymbol{LFI:FWHM:uncertainty:LFI20}{0.170}
\setsymbol{LFI:FWHM:uncertainty:LFI21}{0.156}
\setsymbol{LFI:FWHM:uncertainty:LFI22}{0.164}
\setsymbol{LFI:FWHM:uncertainty:LFI23}{0.171}
\setsymbol{LFI:FWHM:uncertainty:LFI24}{0.652}
\setsymbol{LFI:FWHM:uncertainty:LFI25}{1.075}
\setsymbol{LFI:FWHM:uncertainty:LFI26}{1.131}
\setsymbol{LFI:FWHM:uncertainty:LFI27}{1.266}
\setsymbol{LFI:FWHM:uncertainty:LFI28}{1.287}

\setsymbol{HFI:center:frequency:100GHz:units}{100\,GHz}
\setsymbol{HFI:center:frequency:143GHz:units}{143\,GHz}
\setsymbol{HFI:center:frequency:217GHz:units}{217\,GHz}
\setsymbol{HFI:center:frequency:353GHz:units}{353\,GHz}
\setsymbol{HFI:center:frequency:545GHz:units}{545\,GHz}
\setsymbol{HFI:center:frequency:857GHz:units}{857\,GHz}

\setsymbol{HFI:center:frequency:100GHz}{100}
\setsymbol{HFI:center:frequency:143GHz}{143}
\setsymbol{HFI:center:frequency:217GHz}{217}
\setsymbol{HFI:center:frequency:353GHz}{353}
\setsymbol{HFI:center:frequency:545GHz}{545}
\setsymbol{HFI:center:frequency:857GHz}{857}

\setsymbol{HFI:Ndetectors:100GHz}{8}
\setsymbol{HFI:Ndetectors:143GHz}{11}
\setsymbol{HFI:Ndetectors:217GHz}{12}
\setsymbol{HFI:Ndetectors:353GHz}{12}
\setsymbol{HFI:Ndetectors:545GHz}{3}
\setsymbol{HFI:Ndetectors:857GHz}{4}

\setsymbol{HFI:FWHM:Maps:100GHz:units}{9\parcm88}
\setsymbol{HFI:FWHM:Maps:143GHz:units}{7\parcm18}
\setsymbol{HFI:FWHM:Maps:217GHz:units}{4\parcm87}
\setsymbol{HFI:FWHM:Maps:353GHz:units}{4\parcm65}
\setsymbol{HFI:FWHM:Maps:545GHz:units}{4\parcm72}
\setsymbol{HFI:FWHM:Maps:857GHz:units}{4\parcm39}
\setsymbol{HFI:FWHM:Maps:100GHz}{9.88}
\setsymbol{HFI:FWHM:Maps:143GHz}{7.18}
\setsymbol{HFI:FWHM:Maps:217GHz}{4.87}
\setsymbol{HFI:FWHM:Maps:353GHz}{4.65}
\setsymbol{HFI:FWHM:Maps:545GHz}{4.72}
\setsymbol{HFI:FWHM:Maps:857GHz}{4.39}

\setsymbol{HFI:beam:ellipticity:Maps:100GHz}{1.15}
\setsymbol{HFI:beam:ellipticity:Maps:143GHz}{1.01}
\setsymbol{HFI:beam:ellipticity:Maps:217GHz}{1.06}
\setsymbol{HFI:beam:ellipticity:Maps:353GHz}{1.05}
\setsymbol{HFI:beam:ellipticity:Maps:545GHz}{1.14}
\setsymbol{HFI:beam:ellipticity:Maps:857GHz}{1.19}

\setsymbol{HFI:FWHM:Mars:100GHz:units}{9\parcm37}
\setsymbol{HFI:FWHM:Mars:143GHz:units}{7\parcm04}
\setsymbol{HFI:FWHM:Mars:217GHz:units}{4\parcm68}
\setsymbol{HFI:FWHM:Mars:353GHz:units}{4\parcm43}
\setsymbol{HFI:FWHM:Mars:545GHz:units}{3\parcm80}
\setsymbol{HFI:FWHM:Mars:857GHz:units}{3\parcm67}

\setsymbol{HFI:FWHM:Mars:100GHz}{9.37}
\setsymbol{HFI:FWHM:Mars:143GHz}{7.04}
\setsymbol{HFI:FWHM:Mars:217GHz}{4.68}
\setsymbol{HFI:FWHM:Mars:353GHz}{4.43}
\setsymbol{HFI:FWHM:Mars:545GHz}{3.80}
\setsymbol{HFI:FWHM:Mars:857GHz}{3.67}

\setsymbol{HFI:beam:ellipticity:Mars:100GHz}{1.18}
\setsymbol{HFI:beam:ellipticity:Mars:143GHz}{1.03}
\setsymbol{HFI:beam:ellipticity:Mars:217GHz}{1.14}
\setsymbol{HFI:beam:ellipticity:Mars:353GHz}{1.09}
\setsymbol{HFI:beam:ellipticity:Mars:545GHz}{1.25}
\setsymbol{HFI:beam:ellipticity:Mars:857GHz}{1.03}

\setsymbol{HFI:CMB:relative:calibration:100GHz}{$\lsim 1\%$}
\setsymbol{HFI:CMB:relative:calibration:143GHz}{$\lsim 1\%$}
\setsymbol{HFI:CMB:relative:calibration:217GHz}{$\lsim 1\%$}
\setsymbol{HFI:CMB:relative:calibration:353GHz}{$\lsim 1\%$}
\setsymbol{HFI:CMB:relative:calibration:545GHz}{}
\setsymbol{HFI:CMB:relative:calibration:857GHz}{}

\setsymbol{HFI:CMB:absolute:calibration:100GHz}{$\lsim 2\%$}
\setsymbol{HFI:CMB:absolute:calibration:143GHz}{$\lsim 2\%$}
\setsymbol{HFI:CMB:absolute:calibration:217GHz}{$\lsim 2\%$}
\setsymbol{HFI:CMB:absolute:calibration:353GHz}{$\lsim 2\%$}
\setsymbol{HFI:CMB:absolute:calibration:545GHz}{}
\setsymbol{HFI:CMB:absolute:calibration:857GHz}{}

\setsymbol{HFI:FIRAS:gain:calibration:accuracy:statistical:100GHz}{}
\setsymbol{HFI:FIRAS:gain:calibration:accuracy:statistical:143GHz}{}
\setsymbol{HFI:FIRAS:gain:calibration:accuracy:statistical:217GHz}{}
\setsymbol{HFI:FIRAS:gain:calibration:accuracy:statistical:353GHz}{2.5\%}
\setsymbol{HFI:FIRAS:gain:calibration:accuracy:statistical:545GHz}{1\%}
\setsymbol{HFI:FIRAS:gain:calibration:accuracy:statistical:857GHz}{0.5\%}

\setsymbol{HFI:FIRAS:gain:calibration:accuracy:systematic:100GHz}{}
\setsymbol{HFI:FIRAS:gain:calibration:accuracy:systematic:143GHz}{}
\setsymbol{HFI:FIRAS:gain:calibration:accuracy:systematic:217GHz}{}
\setsymbol{HFI:FIRAS:gain:calibration:accuracy:systematic:353GHz}{}
\setsymbol{HFI:FIRAS:gain:calibration:accuracy:systematic:545GHz}{7\%}
\setsymbol{HFI:FIRAS:gain:calibration:accuracy:systematic:857GHz}{7\%}

\setsymbol{HFI:FIRAS:zero:point:accuracy:100GHz:units}{0.8\MJysr}
\setsymbol{HFI:FIRAS:zero:point:accuracy:143GHz:units}{}
\setsymbol{HFI:FIRAS:zero:point:accuracy:217GHz:units}{}
\setsymbol{HFI:FIRAS:zero:point:accuracy:353GHz:units}{1.4\MJysr}
\setsymbol{HFI:FIRAS:zero:point:accuracy:545GHz:units}{2.2\MJysr}
\setsymbol{HFI:FIRAS:zero:point:accuracy:857GHz:units}{1.7\MJysr}

\setsymbol{HFI:FIRAS:zero:point:accuracy:100GHz}{0.8}
\setsymbol{HFI:FIRAS:zero:point:accuracy:143GHz}{}
\setsymbol{HFI:FIRAS:zero:point:accuracy:217GHz}{}
\setsymbol{HFI:FIRAS:zero:point:accuracy:353GHz}{1.4}
\setsymbol{HFI:FIRAS:zero:point:accuracy:545GHz}{2.2}
\setsymbol{HFI:FIRAS:zero:point:accuracy:857GHz}{1.7}

\setsymbol{HFI:unit:conversion:100GHz:units}{0.2415\MJysrmK}
\setsymbol{HFI:unit:conversion:143GHz:units}{0.3694\MJysrmK}
\setsymbol{HFI:unit:conversion:217GHz:units}{0.4811\MJysrmK}
\setsymbol{HFI:unit:conversion:353GHz:units}{0.2883\MJysrmK}
\setsymbol{HFI:unit:conversion:545GHz:units}{0.05826\MJysrmK}
\setsymbol{HFI:unit:conversion:857GHz:units}{0.002238\MJysrmK}

\setsymbol{HFI:unit:conversion:100GHz}{0.2415}
\setsymbol{HFI:unit:conversion:143GHz}{0.3694}
\setsymbol{HFI:unit:conversion:217GHz}{0.4811}
\setsymbol{HFI:unit:conversion:353GHz}{0.2883}
\setsymbol{HFI:unit:conversion:545GHz}{0.05826}
\setsymbol{HFI:unit:conversion:857GHz}{0.002238}

\setsymbol{HFI:colour:correction:alpha=-2:V1.01:100GHz}{0.9893}
\setsymbol{HFI:colour:correction:alpha=-2:V1.01:143GHz}{0.9759}
\setsymbol{HFI:colour:correction:alpha=-2:V1.01:217GHz}{1.0007}
\setsymbol{HFI:colour:correction:alpha=-2:V1.01:353GHz}{1.0028}
\setsymbol{HFI:colour:correction:alpha=-2:V1.01:545GHz}{1.0019}
\setsymbol{HFI:colour:correction:alpha=-2:V1.01:857GHz}{0.9889}

\setsymbol{HFI:colour:correction:alpha=0:V1.01:100GHz}{1.0008}
\setsymbol{HFI:colour:correction:alpha=0:V1.01:143GHz}{1.0148}
\setsymbol{HFI:colour:correction:alpha=0:V1.01:217GHz}{0.9909}
\setsymbol{HFI:colour:correction:alpha=0:V1.01:353GHz}{0.9888}
\setsymbol{HFI:colour:correction:alpha=0:V1.01:545GHz}{0.9878}
\setsymbol{HFI:colour:correction:alpha=0:V1.01:857GHz}{1.0014}

\providecommand{\sorthelp}[1]{}

\def\eg{e.g.\/}

\def\kms{{\rm km\,s}$^{-1}$}

\def\L{\emph{(Left)}}

\newlength{\thsize}
\newlength{\hhsize}
\newlength{\qhsize}
\def\Planck{\textit{Planck}}

\begin{document}

\title{Dust models post-Planck: constraining the far-infrared opacity of dust in the diffuse interstellar medium}


   \author{L. Fanciullo, \inst{1}
	V. Guillet, \inst{1}
	G. Aniano, \inst{1}
	A. P. Jones, \inst{1}
         N. Ysard, \inst{1}
         M.-A. Miville-Desch\^enes, \inst{1}
	F. Boulanger, \inst{1}
         \and
	M. K\"ohler \inst{1}
          }


   \institute{Institut d'Astrophysique Spatiale (IAS), B\^atiment 121, Universit\'e Paris-Sud 11 and CNRS, F- 91405 Orsay, France}

   \date{}

 
  \abstract
 {}
{We compare the performance of several dust models in reproducing the dust spectral energy distribution (SED) per unit extinction in the diffuse interstellar medium (ISM). We use our results to constrain the variability of the optical properties of big grains in the diffuse ISM, as published by the \Planck\ collaboration.}
 {We use two different techniques to compare the predictions of dust models to data from the \Planck\ HFI, IRAS and SDSS surveys. First, we fit the far-infrared emission spectrum to recover the dust extinction and the intensity of the interstellar radiation field (ISRF). Second, we infer the ISRF intensity from the total power emitted by dust per unit extinction, and then predict the emission spectrum. In both cases, we test the ability of the models to reproduce dust emission and extinction at the same time. }
 {We identify two issues. Not all models can reproduce the average dust emission per unit extinction: there are differences of up to a factor $\sim2$ between models, and the best accord between model and observation is obtained with the more emissive grains derived from recent laboratory data on silicates and amorphous carbons. All models fail to reproduce the variations in the emission per unit extinction if the only variable parameter is the ISRF intensity: this confirms that the optical properties of dust are indeed variable in the diffuse ISM.}
 {Diffuse ISM observations are consistent with a scenario where both ISRF intensity and dust optical properties vary. The ratio of the far-infrared opacity to the $V$ band extinction cross-section presents variations of the order of $\sim20\%$ ($40-50\%$ in extreme cases), while ISRF intensity varies by $\sim30\%$ ($\sim60\%$ in extreme cases). This must be accounted for in future modelling. }

\def\Gex{\gamma_{\rm V}}
\def\Gem{\gamma_{\rm 353}}
\def\sGex{\sigma(\gamma_{\rm V})}
\def\sGem{\sigma(\gamma_{\rm 353})}
\def\Rv{R_{\rm V}}
\def\Av{A_{\rm V}}
\def\Avobs{\Av^{\rm obs}}
\def\Avfit{\Av^{\rm fit}}
\def\Gfit{G_0^{\rm fit}}
\def\Gr{G_0^\mathcal{R}}
\def\Umin{U_{\rm min}}
\def\EBV{E(B-V)}
\def\kfir{\kappa_{\rm FIR}}
\def\tfir{\tau_{\rm FIR}}
\def\kv{\kappa_{\rm V}}
\def\tv{\tau_{\rm V}}

\def\Gonzalofactor{1.9}

\def\fmax{f_{\rm max}}
\def\stiff{s}
\def\athresh{a_{\rm alg}}
\def\Cext{C_{\rm ext}}
\def\Cabs{C_{\rm abs}}
\def\Csca{C_{\rm sca}}
\def\Ca{C_{\rm a}}
\def\Cb{C_{\rm b}}
\def\Cavg{C_{\rm avg}}

\authorrunning{L. Fanciullo et al.}

\titlerunning{Dust models post-Planck}

\maketitle

\section{Introduction}
\label{sect_introduction}

Interstellar dust is an important component of the interstellar medium (ISM): it is studied for its role in the physics and chemistry of the ISM \citep{grainchem,Mathis_review,grainphoto,grainH2}, for its role as a tracer of gas column density \citep{Boul96,Avmap,P06B} and of magnetic field structure \citep{Chap_MF,Poidevin_pol,Berd2014,PIP75_dummy}, and for its effect as a foreground in studies of the cosmic microwave background \citep{PIP87, Planck_pol}. Many dust models have been made to reproduce the main dust observables, which include extinction curve and albedo, spectral energy distribution (SED) from the near-infrared to the microwave continuum, and elemental abundance constraints \citep{ZDw04,DL07,C11,J13}. 

Our understanding of interstellar dust, however, is still incomplete. One of the issues faced by the models is that dust is not the same everywhere in the ISM. As revealed by numerous recent observations \citep{Pagani,Koehler_11,PEP_XXIV,PEP_XXV,Koehler_12,Martin12,Paradis,Y13}, dust properties vary from the diffuse ISM to molecular clouds. Variations are especially notable in the far-infrared opacity, i.e. the far-infrared optical depth per unit dust column density $\tau/N_H$. The dust optical depth $\tau_\lambda$ is the function that, in the optically thin limit, modulates the SED: $I_\lambda = \tau_\lambda \cdot B_\lambda(T)$, where $B_\lambda(T)$ is the black-body emission at a temperature $T$
\footnote{Quantities that depend on dust optical properties, such as optical depth or opacity, are wavelength-dependent. When, in this article, we mention these quantities without specifying a wavelength, the quantities should be interpreted as an ``effective'' value integrated over the whole wavelength range of interest. For instance, since we are interested in the far-infrared emission of grains, we often mention the ``far-infrared opacity'': this is the integral of the opacity over the far-infrared and submillimeter wavelength range, weighted by the emission spectrum.}.

Variations in dust far-infrared opacity can now be traced in the diffuse ISM itself thanks to the survey of the \Planck\ satellite \citep{Survey_Planck}. The far-infrared opacity shows significant variations in the diffuse ISM, although those variations are smaller than the variations observed between the diffuse ISM and molecular clouds \citep{Bot,PIP82,P06B}. 

\cite{Gonzalo} examines the dust emission obtained from \Planck\ , IRAS \citep{survey_IRAS}  and WISE \citep{survey_WISE} data, fitting it over the whole sky with the \cite{DL07} dust model, which uses constant optical properties
\footnote{The optical properties for the materials used in \cite{DL07} were modified as explained in \cite{AD12}.}
. In this paper, our analysis concentrates on more than 200,000 lines of sight towards quasi-stellar objects (QSOs) observed in the Sloan Digital Sky Survey \citep[SDSS,][]{Survey_SDSS} and for which $\Av$ is available 
\footnote{The SDSS survey actually provides the reddening ${\rm E(B-V)}$, from which $\Av$ is computed assuming $\Av/{\rm E(B-V)} = 3.1$. Throughout the present paper, as in \cite{Gonzalo}, whenever the observed $\Av$ is mentioned, the measure referred to is actually $3.1~\cdot~{\rm E(B-V)}$.}
: the $\Av$ from the QSOs is correlated with the value $\Avfit$ deduced from the \Planck\ and IRAS fit.
\cite{Gonzalo} find that Galactic dust SEDs can be well fitted by the Draine \& Li model \citep{DL07} when one uses as free parameters the dust column density -- or, equivalently, the $V$ band extinction $\Av$ -- and the intensity $\Umin$ of the interstellar radiation field (ISRF) in the diffuse ISM. However, even in the diffuse ISM, this model cannot explain the variations in the SED per unit extinction if the ISRF intensity is the only quantity that varies
\footnote{The advantage of normalising dust emission to $\Av$ instead of the HI column density is that the data contain information on dust properties alone, without the uncertainty introduced by the dust-to-gas mass ratio.} 
: the \cite{DL07} model can reproduce the dust SED, but it systematically overestimates the observed $\Av$ by a factor  that depends on $\Umin$ and has a mean value of $\Gonzalofactor$ in the diffuse ISM.

We aim to extend the analysis by \cite{Gonzalo} to the \cite{C11} and \cite{J13} dust models and to quantify variations of their far-infrared emission properties in the diffuse ISM. Our analysis will focus on the big grains in thermal equilibrium (radius $\sim 100$ nm).
The paper is organized as follows. In Sect.~\ref{sect_data_methodo} we introduce the observational data and the dust models used to fit these data. In Sect.~\ref{sect_results} we 
present two different methodologies for constraining dust parameters using observations, we show how their results differ and we estimate the variation of dust optical properties in the diffuse ISM. In Sect.~\ref{sect_discussion} we discuss the results, identifying a possible shortcoming of most current dust models. Finally, in Sect.~\ref{sect_conclusion_and_perspectives} we explore the prospects for future dust models.

\section{Data and models}
\label{sect_data_methodo}

\subsection{Data}
\label{sect_data}
\cite{Gonzalo} provides a family of 20 $A_V$-normalised SEDs, $I_\lambda / \Av$, to be used as constraints for future dust models. The quantity $\Av$ is measured on QSO lines-of-sight: each SED is the average over $\sim$ 10,000 lines of sight with similar $\Umin$. The SEDs have $\Umin$ in a range of 0.42 to 0.98, with an average of 0.66 and a standard deviation of 0.14. The QSO lines of sight sample regions of low $\Av$ with a median value of $\sim0.10$, however, the set of SEDs describe the data over a much larger fraction of the sky ($\ge 70\%$) with $\Av$ up to 1
\footnote{G.~Aniano, private communication.}. The wavelengths coverage is from 60~$\mu$m to 2.1~mm (between 5000 and 143 \GHz).

We use a selection from the \cite{Gonzalo} SEDs. Dust models agree in attributing the $V$ band extinction mainly to big grains, which, being in thermal equilibrium, also dominate emission at $\lambda > 100~\mu$m. To compare the same grain type in extinction and in emission, therefore, we use the 100~$\mu$m (3000 \GHz) IRAS band and the 350, 550, and 850~$\mu$m (857, 545 and 353 \GHz) \Planck\ bands. We exclude the 60~$\mu$m band, where emission is dominated by stochastically-heated smaller grains, because a single band is not sufficient to test small grains properties: a measure of the extinction in the UV is needed to determine the total mass of small particles, and a full mid- and near-infrared emission spectrum is needed to constrain the small particle temperature and size distribution.

\begin{figure}
\includegraphics[width=\hsize]{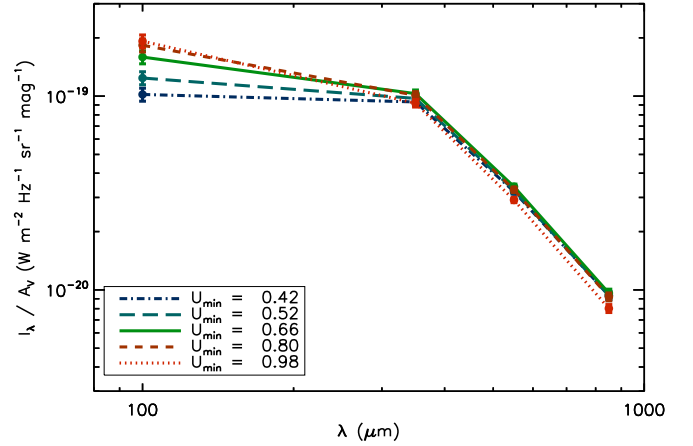}
\caption{
Subset from the 20 $\Av$-normalised dust SEDs in \cite{Gonzalo}. Each $\Umin$, i.e. each temperature, is shown in a different colour and line style. For clarity of illustration, only five of the 20 SEDs are shown: the warmest and coldest overall SEDs, the warmest and coldest of the SEDs that are within a standard deviation from the average $\Umin$, and the median SED. Symbols indicate the central band wavelengths; error bars indicate the dispersion for each band flux, while instrumental noise is negligible.
} 
\label{figure_SEDs}
\end{figure}

Fig.~\ref{figure_SEDs} shows a subset of the data: the median SED, with $\Umin = 0.66$; the two extreme SEDs, with $\Umin = 0.98$ and $\Umin = 0.42$; the SEDs closest to being $1~\sigma$ above and below the median, with $\Umin = 0.80$ and 0.52 respectively. The figure shows that the $\Av$-normalised SEDs are very similar over the Planck wavelength range, and most of the variations are in the IRAS 100~$\mu$m  band.

The SEDs have both statistical and systematic uncertainties. What is important in statistical uncertainties is their relative value from one band to another, which defines the statistical weight to be applied to each band in the fitting routine. The statistical errors in each band of $I_\lambda / \Av$ are negligible, since each SED is an average over more than $10,000$ observations.
The systematic uncertainties in the dust SED come mainly from the photometric calibration: the calibration uncertainty is important at 100~$\mu$m  ($1 \sigma$ error bar = $13.6 \%$) and at 350 and 550~$\mu$m ($1 \sigma$ = $10 \%$); we decided to neglect it at longer wavelengths where it is much smaller ($\lessapprox 1 \%$) \citep{IRIS,Planck_cal}. We estimated the effect of this uncertainty via Monte Carlo simulations. Specifically, for each observed SED, we realized 1000 simulations of random Gaussian-distributed errors for the 100, 350, and 550 $\mu$m bands, with the photometric error at 350 and 550 $\mu$m set equal since both channels are calibrated on planets (the uncertainties are dominated by the planet model used, not by noise). We performed a fit on all simulations, obtaining 1000 values of each fit parameter for each SED: the mean of these was taken as the fiducial value of the fit parameter and the standard deviation as the uncertainty. The systematic uncertainty in $\Av$ is estimated by \cite{Gonzalo} to be $\sim 15\%$.

\subsection{Dust models}
\label{sect_dust_models}

We use three dust models with different characteristics: \cite{DL07}, with the modifications explained in \cite{AD12}, \cite{C11}, and \cite{J13}.

\textbf{\cite{DL07}}, hereafter \textbf{DL07}, is one of the most popular dust models, often used to estimate dust and gas masses in Galactic and extragalactic environments by fitting dust emission \citep{DraineSINGS07,AD12}. It comprises a size distribution of big silicate grains with optical properties adapted from observations \citep{DL84,LiDraine01} and carbonaceous grains whose optical properties are assumed to vary continuously from those of graphite for large sizes ($\gg 10^5$ C atoms) to those of polycyclic aromatic hydrocarbons (PAHs) for small sizes ($\lessapprox 10^5$ C atoms). 

\textbf{\cite{C11}}, hereafter \textbf{C11}, uses silicate grains with the same properties as DL07, but carbonaceous grains in C11 are divided into two populations: one of PAHs, and one population of amorphous carbon grains, covering a size range from nm-sized to big grains, with optical properties derived from \cite{Zubko_96}.

\textbf{\cite{J13}}, hereafter \textbf{J13}, is a core-mantle model with optical properties that are derived from laboratory measurements. The J13 model features big grains of amorphous forsterite-like silicates, with iron inclusions and amorphous aromatic carbon mantles; big amorphous carbon grains with aliphatic cores and aromatic mantles; and small aromatic grains (down to sub-nm sizes), which take the place of PAHs. A unique feature of J13 is that it allows for a certain flexibility in its optical properties, which vary according to the band gap of carbonaceous materials and the size of small grains, as well as possible variations in the mantle thickness. We use the standard model aromatic carbon with a band gap of 0.1 eV.

The dust extinction and emission are computed by interpolating an existing library in the case of DL07 \citep{AD12} and using the DustEM tool
\footnote{http://www.ias.u-psud.fr/DUSTEM/}
 in the case of C11 and J13 \citep{C11}. The observational SEDs were fitted using a $\chi^2$-minimising procedure. We assumed an interstellar radiation field (ISRF) of the form $u_\lambda = G_0 \times u_\lambda^{\rm MMP83}$, where $u_\lambda^{\rm MMP83}$ is the mean ISRF in the solar neighbourhood estimated by \cite{Mathis} and $G_0$ is a dimensionless scale factor. The factor $G_0$ has the same meaning in the model as the $\Umin$ introduced in Sect.~\ref{sect_introduction}, but we chose to use a different name to more easily distinguish our results and those of \cite{Gonzalo}. For the rest of the paper, $G_0$ is our estimate of the ISRF intensity, while $\Umin$ is only used to indicate the SEDs and is functionally synonymous with dust temperature.

\section{Methodology and results}
\label{sect_results}

We analysed the observations using two different methods. In the first method, we fit the SED to obtain the model dust parameters: the radiance, i.e. the integrated SED intensity, the ISRF intensity $G_0$, and the dust column density, as represented by $\Av$. In the second method, we estimate the ISRF intensity from the observed $\Av$ and dust radiance, and then use it to predict the model dependent SED.

\subsection{SED fitting}
\label{sect_SED}

\begin{figure}
\includegraphics[width=\hsize]{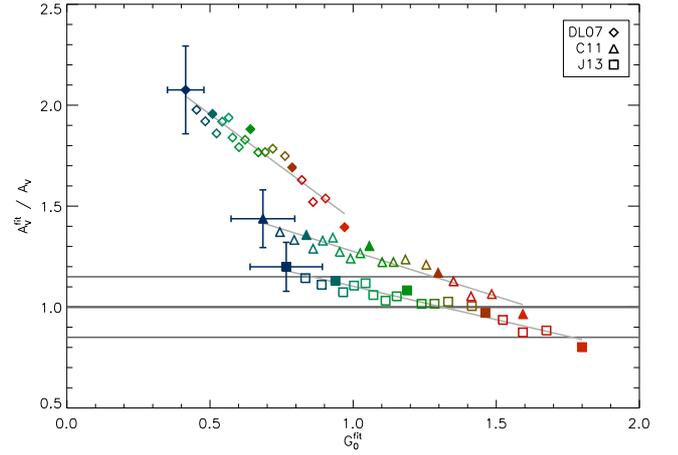}
\caption{Plot of the fitted parameters $\Gfit$ and $\Avfit/\Av$, for the three models. The colour scale goes from blue (cold) to red (warm). The filled symbols correspond to the SEDs shown in fig. \ref{figure_SEDs}. All points have similar relative uncertainty; the error bars for the lowest $\Umin$ are shown. The uncertainties are systematic, so they affect each point in the same way; errors on the two axes are strongly anti-correlated \emph{(see Sect.~\ref{sect_SED})}. The horizontal grey lines show $\Avfit/\Av = 1$ and the $15\%$ uncertainty in the $\Av$ normalisation (Sect.~\ref{sect_data}). The slanted grey lines are the least-square fits for each model; they were added to more easily tell apart the three models.
}
\label{figure_fits_phys}
\end{figure}

The ISRF intensity $\Gfit$ is obtained from the fit of the SEDs. The same fit, since the SEDs are normalised by the QSO-derived $\Av$, returns the ratio of  the model extinction to the observed extinction, $\Avfit/\Av$. The fit results are shown in Fig.~\ref{figure_fits_phys}; each symbol represents a pair of values ($\Gfit$, $\Avfit / \Av$) for a different model and SED. A perfect model would find a value of $\Avfit / \Av = 1$ (thick horizontal grey line). The thinner horizontal grey lines show our $15\%$ systematic uncertainty in $\Av$ (see Section \ref{sect_data}). This uncertainty is identical for all SEDs, and so it affects the significance of the average $\Avfit/\Av$ of the models, but does not affect either the dependence of $\Avfit / \Av$ on $G_0$ or the differences between models. The error bars on ($\Gfit$, $\Avfit / \Av$) are obtained from Monte-Carlo simulations, as explained in Sect.~\ref{sect_data}. They are strongly anti-correlated, with an average Pearson correlation coefficient of $-0.77$ (DL07) to $-0.84$ (C11 and J13). Since these uncertainties are systematic, the plots may be shifted vertically or horizontally, but their shape remains essentially the same. 

The J13 model fits the data well, with an average $\Avfit$ which coincides with the expected value. The C11 model overestimates $\Avfit$ by $\sim25$\%. The DL07 model shows the largest discrepancy, with an $\Avfit$ that is overestimated by a factor $\sim 1.8$, as already pointed out by \cite{Gonzalo}.
All models, however, show a negative correlation between $G_0$ and $\Avfit / \Av$ whereas $\Avfit / \Av$ should be unity
\footnote{While the trends for $\Av$ and $G_0$ in Fig.~\ref{figure_fits_phys} resemble the well-known T-$\tau$ anti-correlation caused by noise \citep{Shetty_II, Shetty_I}, this is not the case here. Since each SED is an average over $10,000$ spectra, the statistical uncertainty for each SED is negligible.}
. For comparison, we also fitted the data with a modified black-body. We obtain similar results, which are presented in Appendix \ref{App-MBB}. 

\subsection{Recovery of $G_0$ from dust radiance}
\label{sect_R}

\begin{figure}
\includegraphics[width=\hsize]{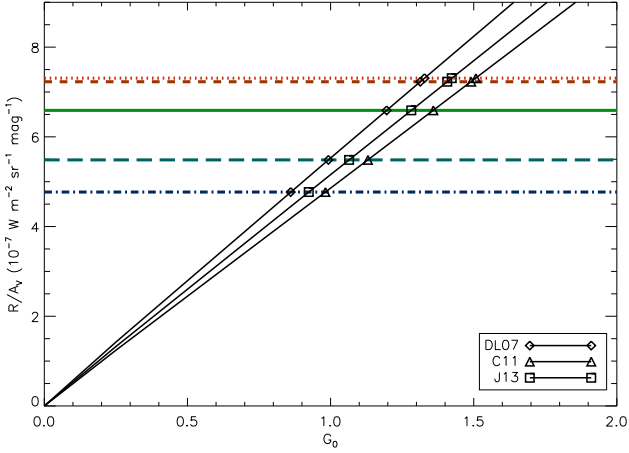}
\caption{Estimation of $G_0$ from the interpolation of $\mathcal{R}/A_{\rm V}$. The horizontal lines are the value of $\mathcal{R}/A_{\rm V}$ for the observed SEDs, the curves from the origin are the $\mathcal{R}^{\rm model}(G_0)/A_{\rm V}^{\rm model}$. The abscissae of the intersection points are the estimates for $G_0$ for that SED and model. The same 5 SEDs as Fig.~\ref{figure_SEDs} are shown, for clarity of illustration, using the same colour and line style scheme.}
\label{figure_G0_from_R}
\end{figure}

The SED fitting yields biased results if the optical properties of the model are not optimal, as was shown by \cite{Gonzalo} for the DL07 model. It would be useful to find an alternative way to compare models and observations. Therefore, we decided, in addition to fitting the dust SEDs, to recover $G_0$ using a procedure based on the dust radiance per unit extinction. 

The radiance, $\mathcal{R}$, is the total power emitted by dust grains in thermal equilibrium \citep{P06B}. Through the conservation of energy, $\mathcal{R}$ corresponds to the power absorbed by grains in thermal equilibrium and can be considered a ``heating power'' for the dust, which is only determined by its absorption properties, independent of its far-infrared opacity.

The thermal emission of big grains can be reasonably well fitted with a modified black-body, $I_\lambda = B_\lambda(T) \cdot \tau_0 \cdot \left(\lambda/\lambda_0\right)^{-\beta}$, with $T$ the observational dust temperature, $\beta$ its spectral index and $\tau_0$ its optical depth at the reference wavelength $\lambda_0$ (here we use 850 $\mu$m, equivalent to a frequency $\nu_0 = 353$ \GHz). The integration of a modified black-body yields \citep{P06B}
\begin{equation} \label{eq_RBB}
\mathcal{R} = \tau_0 \frac{\sigma_s}{\pi} T^4 \left( \frac{k T}{h \nu_0} \right)^{\beta} \frac{\Gamma(4+\beta) \zeta(4+\beta)}{\Gamma(4) \zeta(4)}\,.
\end{equation}
The functions $\Gamma$ and $\zeta$ are the Gamma and Riemann zeta function, respectively, and $\sigma_s$ the Stefan-Boltzmann constant.

For a given dust model one can compute its $\Av$-normalised fluxes in the IRAS and \Planck\ bands, fit a modified black-body to the normalised SED (for $\lambda \geq 100~\mu$m), and substitute the resulting parameters in Eq.~\ref{eq_RBB}. The result, $\mathcal{R}^{\rm model}/A_{\rm V}^{\rm model}$, is a linear function of $G_0$, and can be inverted to find $G_0$ from the radiance of an $\Av$-normalised SED. Fig.~\ref{figure_G0_from_R} shows $\mathcal{R}^{\rm model}(G_0)/A_{\rm V}^{\rm model}$ for the three models. As we mentioned, $\mathcal{R}$ only depends on absorption, and so the estimation of ISRF intensity given by this method, $\Gr$, is not affected by variations in dust opacity. Since the method of obtaining $\Gr$ assumes that the dust optical properties are fixed in extinction, while the method of obtaining $\Gfit$ assumes that optical properties are fixed in both extinction and emission, we expect $\Gr$ to be much less biased than $\Gfit$. This conclusion is supported by the fact that $G_0$ estimates become much more model dependent when they include optical properties in extinction: $\Gr$ varies by $\sim 15\%$ between models, as opposed to the factor $\sim 2$ in $\Gfit$.

\begin{figure*}
\includegraphics[width=0.49\hsize]{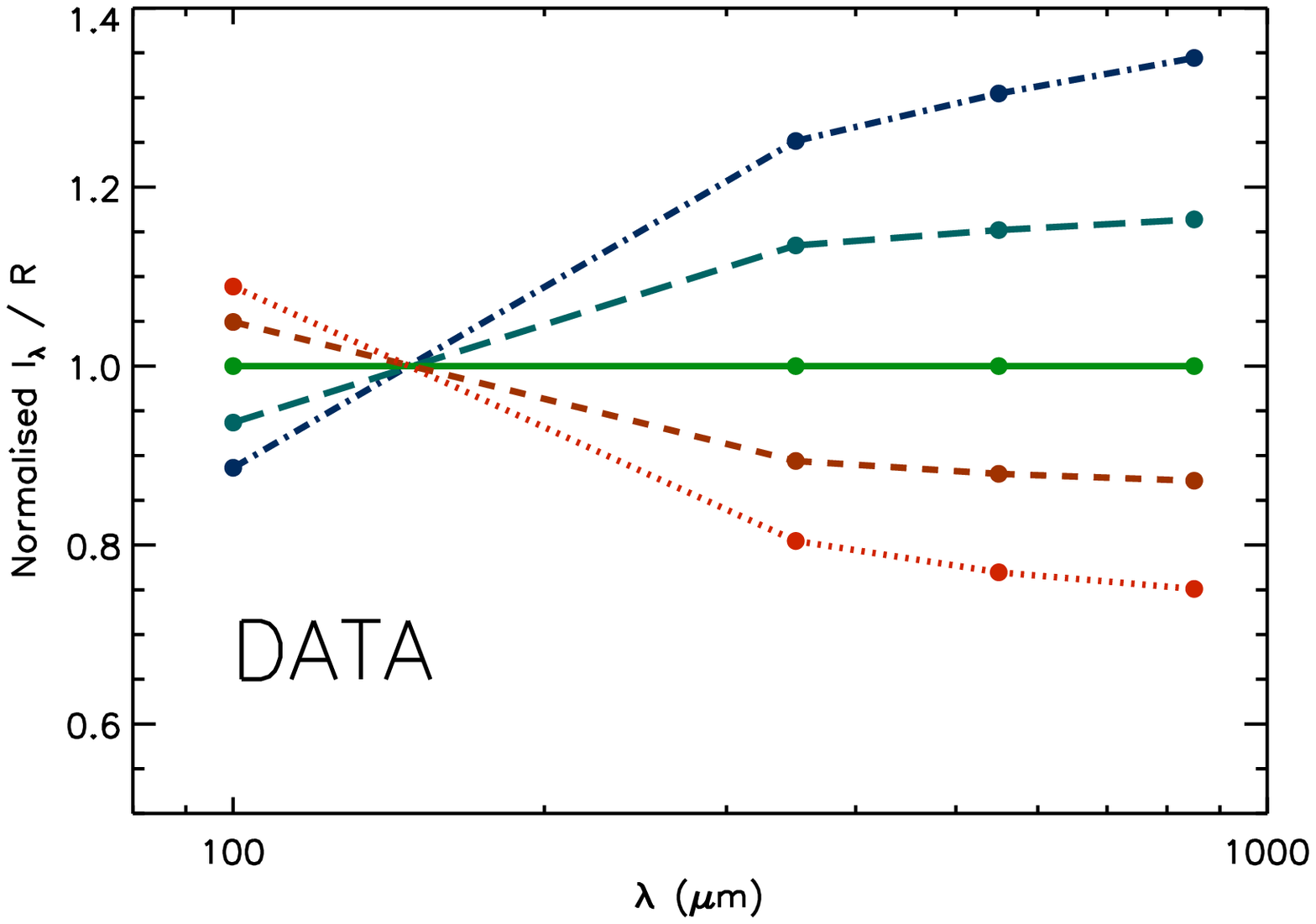}
\includegraphics[width=0.49\hsize]{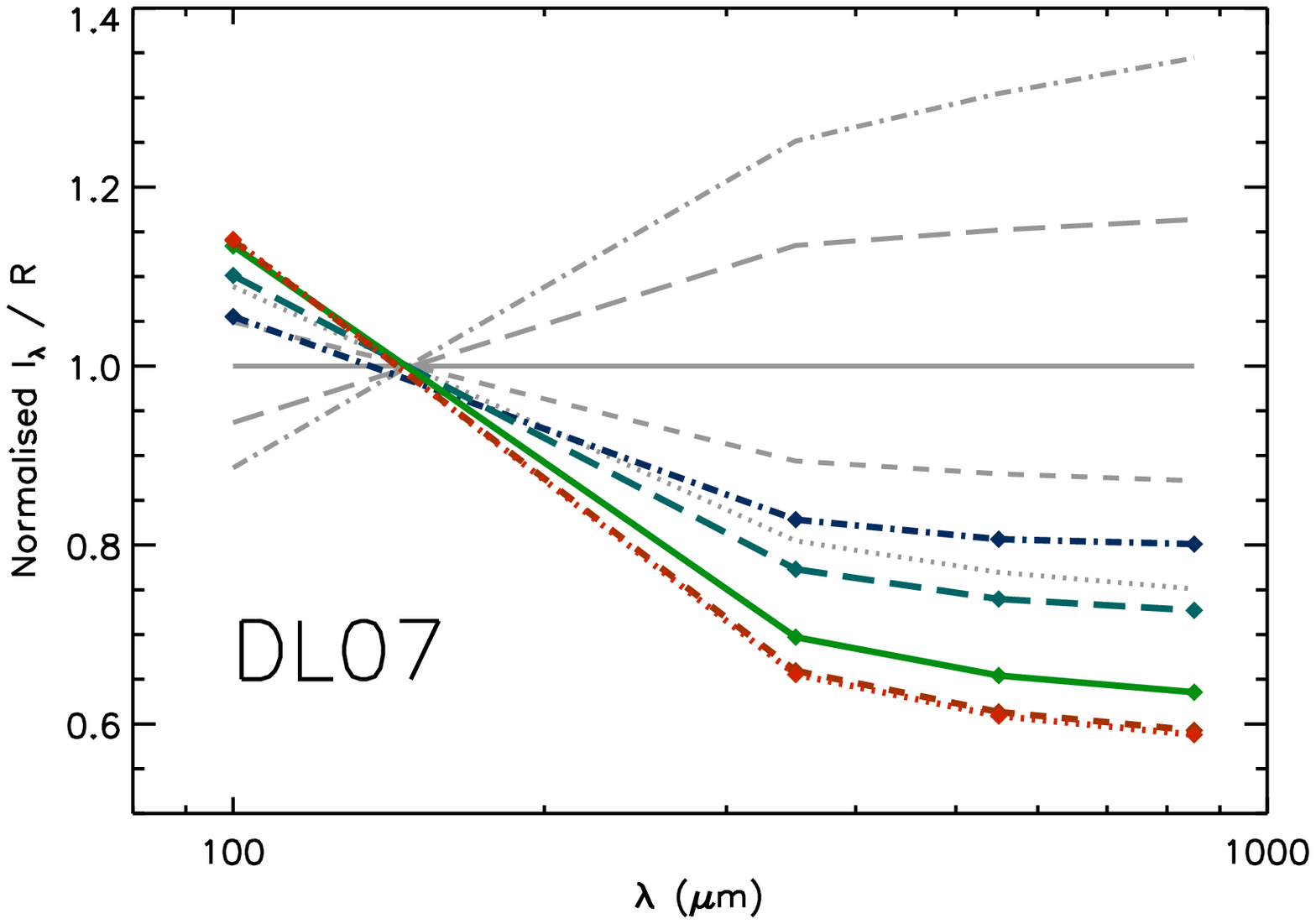}\\
\includegraphics[width=0.49\hsize]{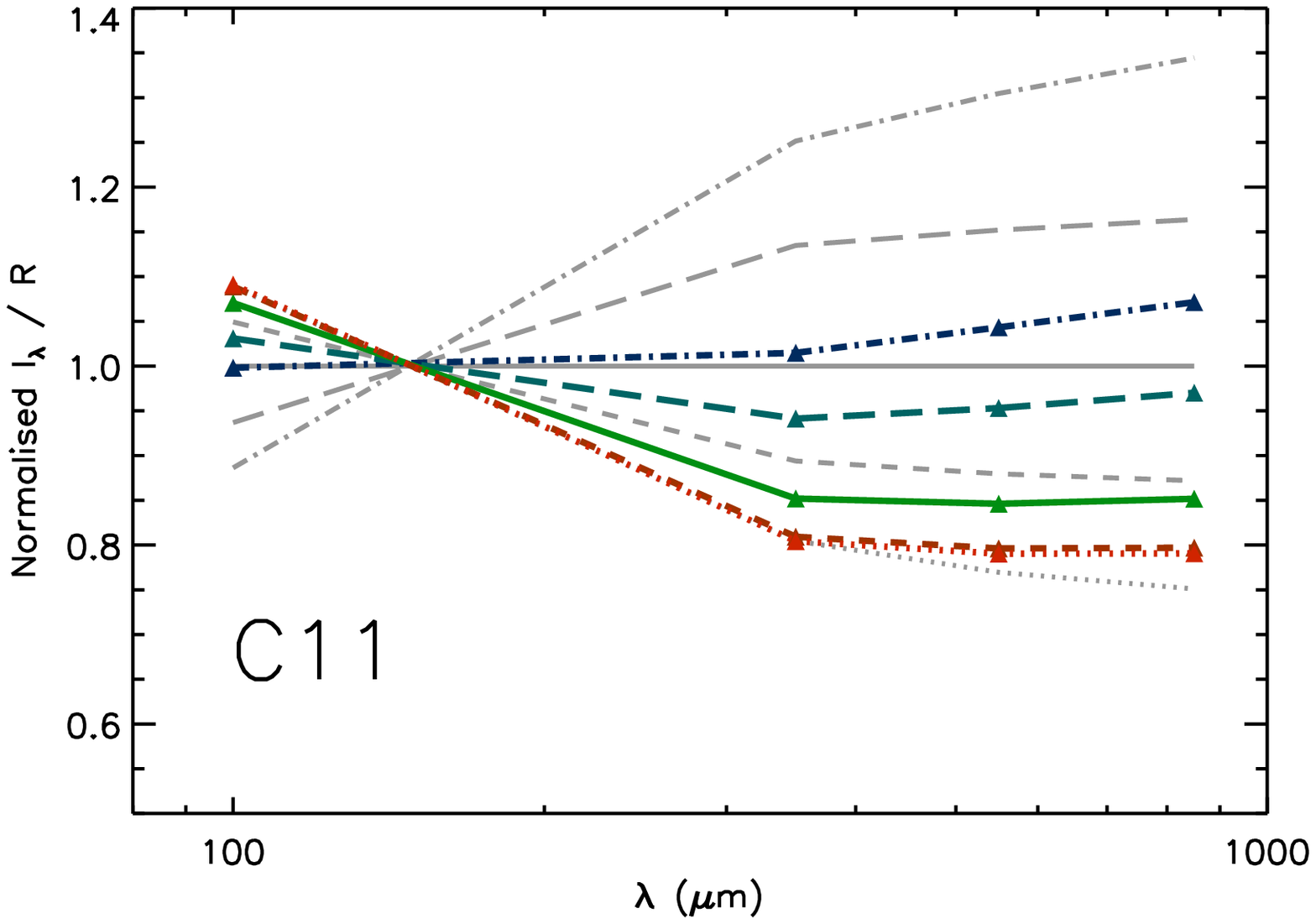}
\includegraphics[width=0.49\hsize]{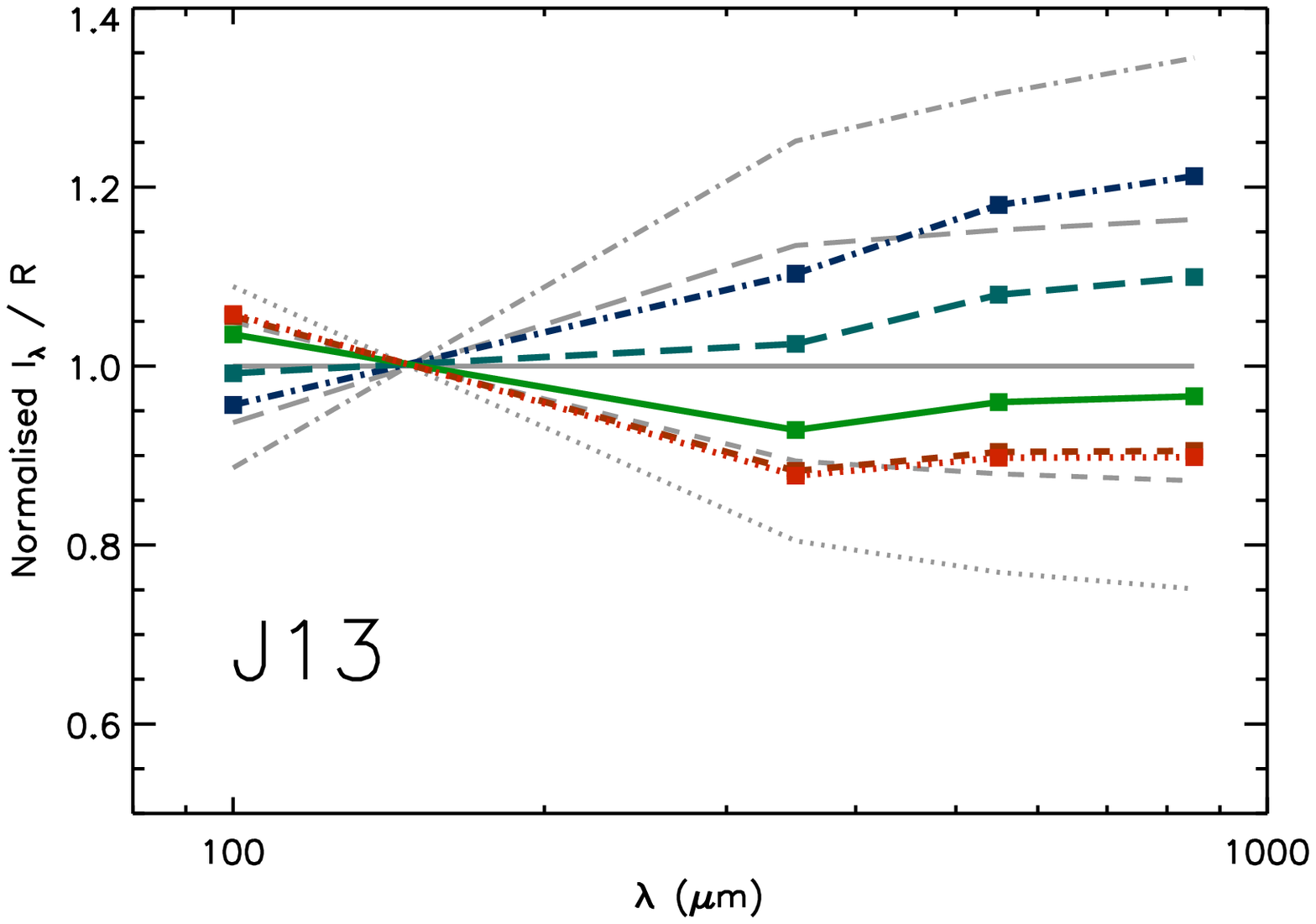}
\caption{Data SEDs (top left) and dust model $I_\lambda/\mathcal{R}$, normalised by the median data SED ($\Umin=0.66$) to highlight the spectral shape variations. The model SEDs are computed using $\Gr$ as ISRF intensity (Sect.~\ref{sect_R}).  The same five SEDs as Fig.~\ref{figure_SEDs} are shown, for clarity of illustration, using the same colour and line style scheme. The observations are plotted in grey behind the models to aid comparison}. 
\label{figure_SED+models_separedR}
\end{figure*}

The values for $\Gr$, shown in Fig.~\ref{figure_G0_from_R}, vary by a factor 1.6 between the two extreme SEDs and by 1.3 if we only consider the SEDs within $1\sigma$ of the average (which would represent the ``typical'' $G_0$ variation rather than the full range). In comparison, in Fig.~\ref{figure_fits_phys}, where the dust optical properties in the model are fixed and the ISRF is the only source of difference between SEDs, $\Gfit$ varies by a factor 2.3 between extremes and 1.6 within the ``$\pm 1 \sigma$'' range.

After obtaining $\Gr$, we use it as a fixed parameter in the computation of the SED of the three dust models. The results are shown in Fig.~\ref{figure_SED+models_separedR}. To highlight spectral shape variations, the SEDs have been normalised by the median data SED and then further divided by their $\mathcal{R}/\Av$. The J13 model reproduces the median SED almost perfectly. Predictions based on the C11 model systematically present an excess emission at short wavelengths and low emission at long wavelengths. This situation is even more drastic in the DL07 model. Furthermore, for all models, the normalised SEDs in Fig.~\ref{figure_SED+models_separedR} spread over a smaller range than the data: the warm (high $\Umin$) SEDs and the cold (low $\Umin$) ones are closer in temperature for the model than they are for the observations. 

The differences shown in Fig.~\ref{figure_SED+models_separedR} between the J13, C11 and DL07 dust models can be attributed to their optical properties in far-infrared emission, as opposed to their optical properties in extinction
\footnote{The models have very similar optical properties in extinction, see Section \ref{sect_R}.}.
We find that the J13 model \citep{J13} is a better model for the diffuse ISM emission than that presented in  \cite{C11} and an even better model than the \cite{DL07} model. The J13 model uses optical properties for silicates and amorphous carbon based on lab measurements. The \cite{C11} and \cite{DL07} models use silicates with far-infrared optical properties extrapolated from mid-infrared astronomical observations \citep{DL84} and empirically adjusted \citep{LiDraine01} to reproduce the FIRAS spectra of \cite{Finkbeiner99}
\footnote{The high-Galactic latitude \cite{Finkbeiner99} spectra, extrapolated from IRAS, predicted a different SED than that subsequently observed by \Planck~\citep{Gonzalo}.}:
these optical properties pertain to silicates that are contaminated by other materials, including possibly a carbonaceous dust component. Furthermore, \cite{DL07} is unique among the models in that its carbon dust is assumed to be graphite.  Apparently, ``astronomical silicate'' and graphite are not emissive enough in the far-infrared and submillimeter.

\subsection{Variation in dust opacity}
\label{sect_variable_oprop}

The models shown in Fig.~\ref{figure_SED+models_separedR} assume constant optical properties and are tuned to reproduce the correct $\mathcal{R}/\Av$ for each SED, meaning that spectral shape variations in the modelled SEDs are solely driven by the ISRF. It is clear that the full range of spectral shapes exhibited by the observations (Fig.~\ref{figure_SED+models_separedR}, top left) cannot be reproduced by ISRF variations alone. As mentioned in Section \ref{sect_R},  if the dust optical properties are fixed the variations of $\Gr$ are only about half of what is needed to reproduce the data. We need dust with variable optical properties to reproduce the observations.

We start by analysing a simplified scenario in which the dust optical properties in extinction are fixed, i.e. $\Av$ is a proxy for the dust column density, and the SED shapes are explained by the variation in far-infrared opacity. Our aim is to quantify this variation, at least to a first approximation. At the end of this section, we see what changes when the assumption of constant extinction is dropped. 

Since the SEDs in our work are normalised by $\Av$, any determination of the far-infrared optical depth from the data actually yields $\tfir / \Av$. In addition, since $\Av$ is a proxy for the dust column density, its variations are identical to those of the opacity. Fig.~\ref{figure_fits_phys} shows that $\Avfit/\Av$ increases by $40-50\%$ between the warmest and coldest SEDs, independent of the model. This indicates that in the diffuse ISM, $\tfir/\Av$ (and therefore opacity) varies by a factor $40-50\%$ between extremes. When considering just the SEDs within $1\sigma$ of the average, we obtain an estimate of the ``typical'' variation: $\sim 20\%$.

\begin{figure*}
\includegraphics[width=0.49\hsize]{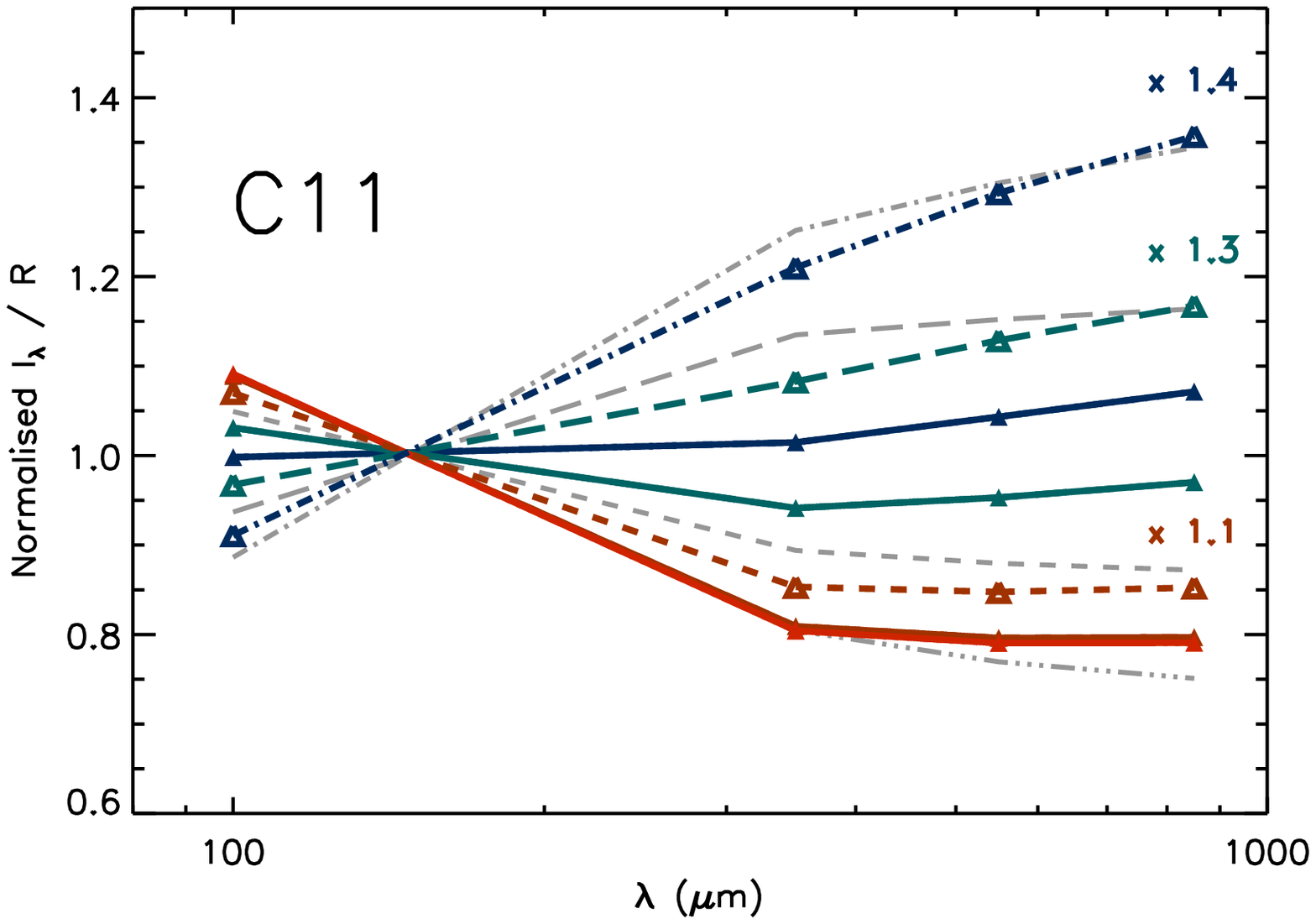}
\includegraphics[width=0.49\hsize]{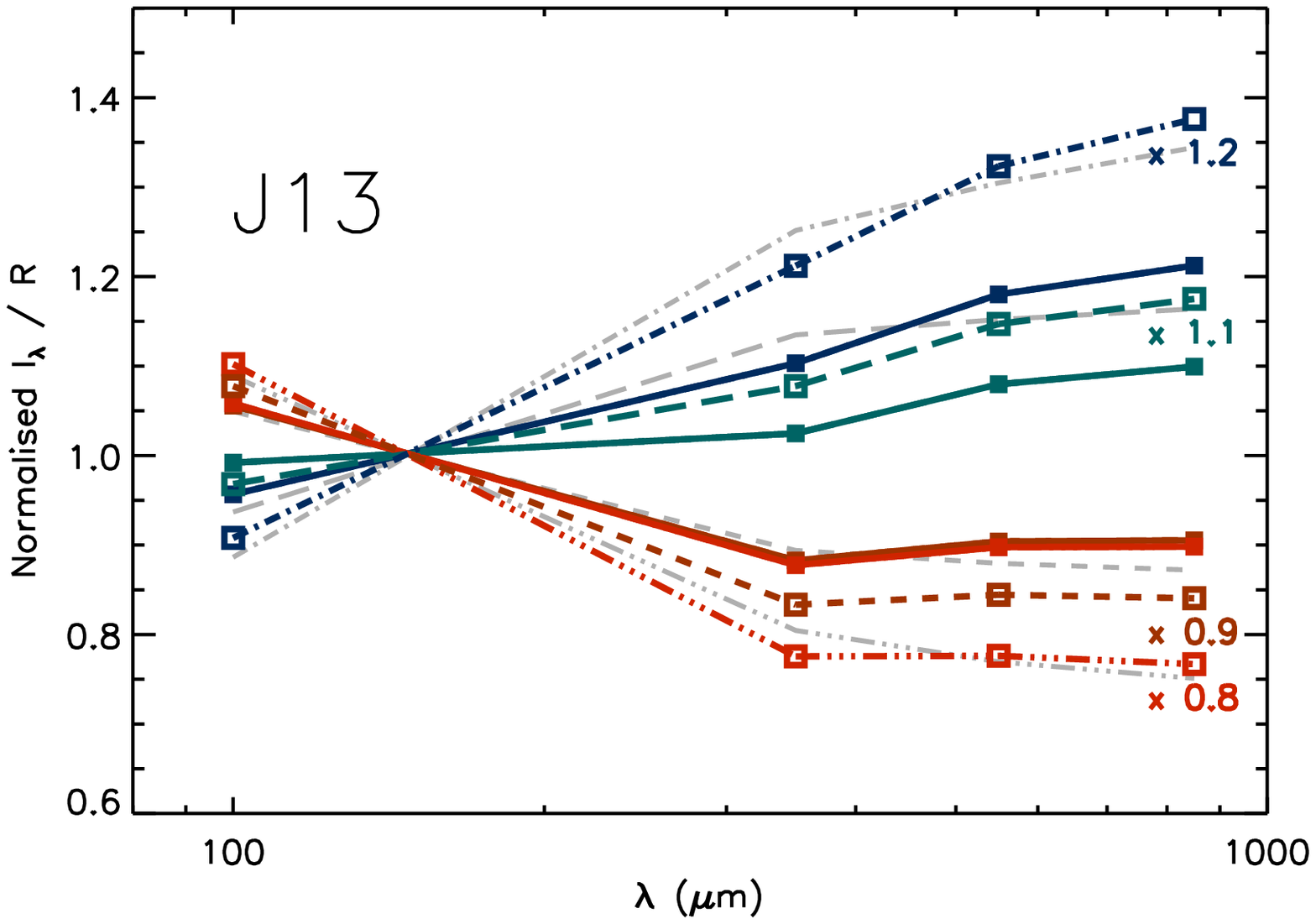}
\caption{Effect of variable far-infrared opacity on $I_\lambda/\mathcal{R}$ for the C11 (left) and J13 (right) models. 
The same five SEDs as Fig.~\ref{figure_SEDs} are shown, except the median SED, using the same colour and line style scheme. The normalisation is the same as Fig.~\ref{figure_SED+models_separedR}. The observations are plotted in grey behind the models; the standard dust models are the solid lines with filled polygonal symbols and the modified dust models are the dashed lines with empty symbols. Numbers show the multiplicative factor of far-infrared opacity. 
}
\label{figure_models_modified}
\end{figure*}

We can compare this estimate to another, obtained with a different technique. We repeat the analysis of Section \ref{sect_R} on modified C11 and J13 models in which the far-infrared opacity of silicates and carbon has been multiplied by the same factor, independent of grain size, as shown in Fig.~\ref{figure_models_modified}. This modification is a purely phenomenological artifice, not based on any physical properties of the materials. We only use this modification to estimate the variations in far-infrared opacity that a physical model should attempt to reproduce. We find that C11 would reproduce all the SEDs within $1\sigma$ of the average if we allowed its far-infrared opacity to vary between 1.1 and 1.3 times its standard value (between 1.0 and 1.4 to reproduce the full range of observations). The results are similar for the J13 model with scaling factors 0.9 and 1.1 for the $1\sigma$ range of SEDs (0.8 and 1.2 for the full range). Thus, we estimate the variations of far-infrared opacity to be $\sim20\%$ ($40 - 50\%$) for the typical (full) range of diffuse ISM. 

So far, we have assumed that optical properties in extinction remain the same while the opacity varies. If we drop the assumption the conclusion must be modified. When the optical properties in extinction vary, $\Av$ is no longer a proxy for the dust column density: our previous conclusions on the variations of $\tfir/\Av$ are still valid, but they no longer apply to opacity either. Since $\Av$ is proportional to the extinction cross-section in the $V$ band, this means that the variation range found above applies to the ratio of the far-infrared opacity to the extinction cross-section in the $V$ band.

\section{Discussion}
\label{sect_discussion}

While some models are closer to the mean SED, no model is able to reproduce the full extent of observed variations in the SEDs. We recall that in the three models the optical properties of dust are assumed constant. Our results show that the variation of $G_0$ alone is not sufficient to reproduce the entire span of spectral shapes presented by the observations. Models that endeavour to explain the emission of the diffuse ISM need to include physical processes that change the far-infrared optical depth per unit $\Av$, or the ratio of the far-infrared opacity to the $V$ band extinction cross-section. 

Variations in size distribution can lead to changes in $\Av$ while having little effect on the far-infrared opacity. However, the resulting variations in $I_{\lambda}/\Av$ are small given the constraints imposed on $G_0$ by $\mathcal{R}/\Av$ (See Appendix~\ref{App-sdist}). A promising route is to consider variations in the grain composition. One process that could alter the grain composition is the accretion of carbon, from the gas phase or from small particles, on big grains. It is readily possible to consider this process within the J13 model because it is tuned to exactly this kind of scenario: its big grains, both silicate and carbonaceous, include an aromatic carbon mantle, the thickness of which can be modified. In the same model, it is also possible to explore the effect of having iron inclusions partially composed of iron sulphide \citep{Koehler_14}. \cite{Y15} shows that the two scenarios, together with variations in the grain size distribution and in the relative abundance of silicate and carbon grains, can explain most of the observed SED variations. Grain shape and structure also play a role in opacity. It is known, for instance, that including porosity in grains increases the far-infrared opacity, while leaving the properties in extinction relatively constant \citep{Ossenkopf_fluffy, Stepnik, Koehler_12}. Porous grains are usually considered typical of the dense ISM rather than the diffuse ISM, but it has been suggested \citep[\eg][]{Martin12} that they could migrate from the dense environment to the diffuse ISM. Far-infrared opacity also increases when one employs non-spherical grains \citep[\eg][]{SVB14}, the existence of which is attested by interstellar polarisation. Determining whether a relation between dust evolution and polarisation exists may be an interesting follow-up analysis of the data. 

\section{Conclusion and perspectives}
\label{sect_conclusion_and_perspectives}

We confront three dust models \citep{DL07,C11,J13} to a family of $\Av$-normalised SEDs obtained from a combination of the \Planck\ , IRAS, and SDSS surveys probing the Galactic diffuse ISM \citep{Gonzalo} .

The widespread method of fitting the SED to obtain the radiation field intensity $G_0$ and the dust column density (from $\Av$) gives results that are very model dependent. The average $\Avfit$ of \cite{J13} is close to the observed value, while that of \cite{C11} is about $25 \%$ too high and that of \cite{DL07} too high by a factor $\sim 1.8$. Furthermore, all models present a non-physical mass-temperature anti-correlation and yield biased $G_0$ estimates.

We develop an alternative $G_0$ estimate that uses the radiance per unit extinction, $\mathcal{R}/A_{\rm V}$. This estimate is robust with respect to variations in the far-infrared optical properties and gives less biased results. We use these values to compute the SEDs for the models, finding that the variations in observed SEDs cannot be reproduced without varying dust optical properties. We estimate that, to reproduce the typical variations observed, we need to vary the ratio of the far-infrared opacity to the $V$ band extinction cross-section by $\sim 20\%$ in the typical diffuse ISM (up to $40-50\%$ to explain the full range of observations); at the same time, variations of $\sim 30\%$ are needed in the ISRF intensity (up to $\sim 60\%$ for the full range of observations).

\begin{acknowledgements}
FB and GA acknowledge support from the ERC grant no. 267934.
\end{acknowledgements}

\bibliographystyle{aa}
\bibliography{Fanciullo_Bibliography}

\appendix
\section{Modified black-body models}\label{App-MBB}

\begin{figure}
\includegraphics[width=\hsize]{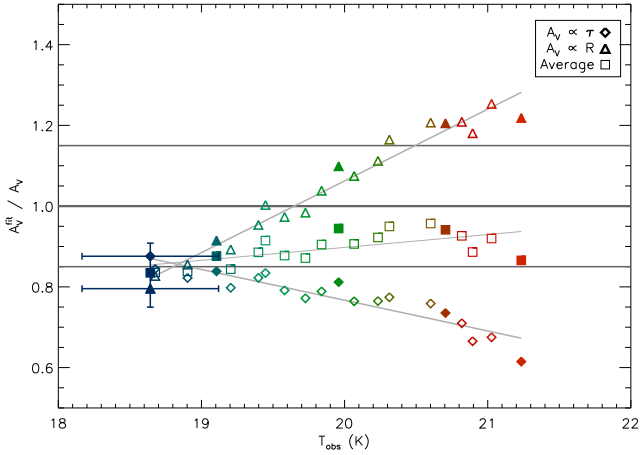}
\caption{Scatter plot of $\Avfit/\Av$ and fitted dust temperature for the modified black-body fit. Two alternative estimates of $\Av$ are shown \emph{(see text)}, plus an average value. 
The horizontal grey lines show $\Avfit/\Av = 1$ and the $15\%$ uncertainty in the $\Av$ normalisation (Sect.~\ref{sect_data}). All points have similar relative uncertainty; the bars for the lowest $U_{min}$ are shown. The uncertainties are systematic, so they affect each point in the same way; errors on the two axes are strongly anti-correlated. The $\Umin$ colour scheme is the same as Fig.~\ref{figure_fits_phys}; the filled symbols correspond to the SEDs shown in fig. \ref{figure_SEDs}. 
}
\label{figure_fits_mbb}
\end{figure}

Fig.~\ref{figure_fits_mbb} shows two different estimates of $\Av$ as a function of temperature, obtained from a modified black-body fit to the observed SEDs. \cite{P06B} provides two different empirical relations for computing the dust reddening $E(B-V)$ from the emission. One relation considers $E(B-V)$ to be proportional to the 850~$\mu$m dust optical depth $(E(B-V)/\tau_0 = 1.49 \cdot 10^4)$, the other relation considers $E(B-V)$ proportional to the radiance $(E(B-V)/\mathcal{R} =  5.4 \cdot 10^5)$. We calculate $E(B-V)$ in both ways, and then we convert the values to $\Av$ using the average diffuse ISM value for $R_V = E(B-V) / A_{\rm V} = 3.1$ and compare the results.

The two estimates differ both by their average value and their trend with temperature: the $\Av$ obtained from $\tau_0$ is about $20 \%$ too low on average and decreases with temperature like the physical models (Fig.~\ref{figure_fits_phys}); the $\Av$ obtained from $\mathcal{R}$ has a good average value but increases with temperature. By way of comparison we also show the geometric average of the two. This average matches the expected value better, despite having no physical justification. Interestingly, the two estimates implicitly make opposite assumptions: the $\Av$ obtained from $\tau_0$ assumes that the dust optical properties are fixed, or at least $\tau_0/A_{\rm V}$ is fixed; and the $\Av$ from $\mathcal{R}$ assumes a fixed $G_0$, or at least a fixed absorbed power per grain
\footnote{The absorbed power depends not only on $G_0$, but also on dust properties such as albedo and size distribution. Our conclusions about variations in $\tau_0$, however, do not change if in the following we consider $G_0$ to mean an ``effective $G_0$'' that includes albedo variations.}
. This means that cold dust is more emissive than expected from models with fixed dust properties, but less emissive than expected from models where variable optical properties account for all observed variations.

\section{Effects of grain size distribution}\label{App-sdist}

The size distribution of dust grains has a strong effect on extinction, but it is not expected to affect the far-infrared opacity, which only depends on the total volume of the grains in the Rayleigh regime. It is natural, therefore, to consider variations in grain size distribution as a way of varying $\tau_{FIR}/\Av$. Assessing the effect of grain size variation in a physically realistic way is not straightforward: the physical processes that change grain sizes, e.g. shattering, sputtering, accretion, coagulation, also affect its structure and composition; also, the optical properties of the materials themselves may be size-dependent. Simply varying the grain size distribution in a model is therefore not likely to mimic the actual variations in the ISM, but can still provide interesting qualitative insights. In this Appendix we explore the effects of varying the size distribution using the C11 model. While this model does not fit the average $I_\lambda/\Av$ as well as J13, it is still close enough to be useful for a differential analysis. Its homogeneous grains and constant optical properties allow us to modify the grain size distribution independent of optical properties. 

In C11, grains larger than $\sim 10$ nm are distributed according to a power law -- $n(a) \propto a^{\alpha}$, where $a$ is the grain radius -- with an exponential cutoff above $\sim150$ nm. The parameter that mainly controls the size distribution is the exponent of the power law, $\alpha$, which is -2.8 for carbonaceous grains and -3.4 for silicate grains. We repeat the procedure of Sect.~\ref{sect_R} varying $\alpha$ by $-0.5$ and $+0.5$ around its standard value. This changes $\Rv$ by $-0.7$ and $+1.0$, respectively; by comparison, \citep{FM07} give $\sim 0.3$ as the typical 1-$\sigma$ dispersion of $\Rv$ in the diffuse ISM.

\begin{figure}
\includegraphics[width=\hsize]{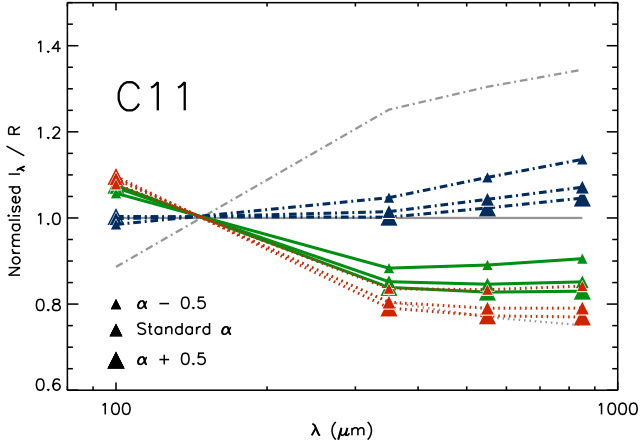}
\caption{
Effect of the grain size distribution on $I_\lambda/\mathcal{R}$, as predicted by the C11 model. 
The SEDs shown are the coldest (blue dot-dashed line), the median (green solid line) and the warmest (red dotted line); normalisation is the same as per Fig.~\ref{figure_SED+models_separedR}. The corresponding observations are plotted in grey behind the models. Larger symbols indicate larger average grain size, and smaller symbols indicate smaller average grain size (see text for details).
}
\label{figure_sdist_effect}
\end{figure}

The results are shown in Fig.~\ref{figure_sdist_effect}. Varying the size distribution has a small impact on the dust SED, and the range of temperatures reproduced is smaller than that observed despite the large, possibly overestimated, span in $\alpha$. The figure also shows that models with smaller grains are, surprisingly, colder than models with larger grains, i.e. they have lower 100-$\mu$m emission and higher long-wavelength emission. This can be explained simply. In our modelling, the radiation field intensity is not fixed, but is rather derived from the observed radiance per unit extinction $\mathcal{R}/\Av$. Changing the size distribution modifies our estimate for $G_0$, so that models with smaller grains necessitate a weaker radiation field to satisfy those constraints. The decrease in $G_0$ thus offsets the temperature increase due to size effects, and even reverses it in the case of the C11 model.

The details of the result presented in this appendix are likely to depend on the dust model and parametrization used. Still, varying the grain size distribution without its corresponding change in the dust optical properties is not likely to explain the observed variations of $I_\lambda/\Av$. 
In \cite{Y15} a similar study uses the J13 model, which is instead adapted to reproduce the interplay of grain size and optical properties.

\end{document}